\documentstyle[12pt]{article}
\def\beq{\begin{equation}}   \def\eeq{\end{equation}}
\def\beqa{\begin{eqnarray}}
\def\eeqa{\end{eqnarray}}

\begin{document}
\title{
\begin{flushright}
{\large\bf IP/BBSR/2000-33\\October, 2000}
\end{flushright}
\vspace{1.5cm}
 {\large\bf Dynamics on $AdS_2$ 
and Enlargement of $SL(2,R)$ to \\
$C = 1$ `cut-off Virasoro Algebra'}} 

\author{{Balram Rai\footnote{E-mail:
balram@iopb.res.in}}\\ Institute of Physics\\
Sachivalaya Marg\\ Bhubaneswar\\ India\,-\,751005}

\date{}
\maketitle

\begin{abstract}
We consider the enhancement of $SL(2,R)$ to Virasoro
algebra in a system of $N$ particles on $AdS_2$.
We restrict our discussion to the case
of non-interacting particles, and argue that they
must be treated as fermions.
We find operators $L_n$
whose commutators on the ground state,
$|vac\rangle$, 
satisfy relations that are reminisent of
$c = 1$ Virasoro algebra, provided
$N \geq n \geq -N$. Same relations hold also on
the states $L_{-k}|vac\rangle\,$, if 
$\,(N-k) \geq n \geq -(N-k)$.
The conditions $L_n^\dag = L_{-n}$, and
$L_k|vac\rangle = 0$ for $k \geq 1$, are also satisfied.
\end{abstract}

\newpage

\subsection*{1. Introduction}

It is of interest to understand the properties of
dynamics in two-dimensional Anti de-Sitter spacetime, $AdS_2$,
for several reasons\cite{st}\cite{gt}\cite{sm}. 
The near horizon geometry of
four-dimensional extremal
black-holes is of the form $AdS_2 \times S_2$. Therefore,
study of dynamics on $AdS_2$ may tell us about the properties
of near horizon degrees of freedom.
Another reason for interest in dynamics on $AdS_2$ is from
the point of view of $AdS_2/CFT_1$ correspondence.
The aim of this paper is to study the dynamics of $N$-Particles
on $AdS_2$ with a view to identifying the symmetries
realised by this system.

First we will consider the case of 
one particle on $AdS_2$. We start by realising the
$SL(2,R)$ symmetry which is to be expeted due to the
fact that the isometry group of $AdS_2$ is $SL(2,R)$.
Surprisingly, the operators $L_1,L_{-1},L_0$ that
we are led to consider satisfy $SL(2,R)$ relations
only on states that are orthogonal to the ground 
state. 
Further we will find that this single particle 
system also realises a semi-infinite 
sub-algebra of Virasoro-algebra. We will obtain operators,
$L_n$, which satisfy the relation
\beq
[\, L_n , L_m \,] = (n-m)\,L_{n+m} \label{introsemi}   
\eeq
where $n,m \geq 0$ or $n,m \leq 0$ .
$L_0$ is the Hamiltonian and $L_n^\dag = L_{-n}$.
        
Next, we will consider $N$-Particles on $AdS_2$.
Any interaction that is introduced between these
particles must be compatible with the $SL(2,R)$
symmetry following from the $AdS_2$ isometry.
It appears complicated to introduce such an interaction
and we will restrict ourselves to non-interacting
system. However, we will argue that we must treat
these particles as fermions. This is seen by imagining
that our system is obtained by gradually turning
off the interaction that is compatible with $SL(2,R)$.
We introduce the $L_n$ operators for multiparticle
system by simply taking the sum of corresponding
single particle
operators, except that now the $L_0$ differs from
the $N$-particle
Hamiltonian by a constant shift.
Obviously, these $N$-particle $L_n$s also satisfy
the semi-infinite Virasoro-algebra of the form 
eqn.(\ref{introsemi}).
The condition $L_n|vac\rangle = 0$,
for $n \geq 1$, is satisfied. The $L_1, L_{-1}, L_0$
satisfy $SL(2,R)$ relations on all states of 
excitation energy less that $N$.
A remarkable fact
that turns out is that, for $N \geq n,m \geq -N$,
\beq
\{\,[\,L_n , L_m\,]\,-\, (n-m)\, L_{n+m} \,-\,
\frac{1}{12} \, (n^3 - n) \delta_{n+m , 0}\,\}\,|vac\rangle
\,=\,0
\eeq
Above equation with $|vac\rangle$ replaced by
$L_{-i}|vac\rangle$ is also valid if   
$(N-i)\geq n,m \geq-(N-i)$.
These relations are reminisent of $c = 1$ Virasoro
algebra, although strictly speaking 
the $L_n$s do not generate Virasoro algebra because
those relations are not satisfied on all the states.
We expect that in general, for a state 
$|\epsilon\rangle$ of excitation
energy $\epsilon$, the relation                              
\beq
\{\,[\,L_n , L_m\,]\,-\, (n-m)\, L_{n+m} \,-\,
\frac{1}{12} \, (n^3 - n) \delta_{n+m , 0}\,\}\,|\epsilon\rangle
\,=\,0
\eeq
will be satisfied if $(N-\epsilon) \geq n,m \geq -(N-\epsilon)$,
and in the large $N$ limit the standard $c=1$ Virasoro structure
will be recovered. 

\subsection*{2. One Particle on $AdS_2$}

The geometry of $AdS_2$ spacetime, of radius of
curvature $R$, is given by the metric
\beq
ds^2 = \frac{R^2}{\sin^2\sigma} \;
( -\, d\tau^2 + d\sigma^2 )  \label{metric}
\eeq
where $-\pi \leq \sigma \leq 0$, and $\tau$ is periodic
with period $2\pi$. The $\sigma$ and $\tau$ are globally
well defined coordinates on $AdS_2$.
Actually, in the following we will work with the
covering space of $AdS_2$ i.e. we consider  
$-\infty < \tau < \infty$, without periodicity.
$AdS_2$ has two boundaries, given by $\sigma = -\pi$
and $\sigma = 0$.
Thus the topology of this spacetime is
that of a strip. The fact that $AdS_2$ can be described
as a hypersurface: $x_0^2-x_1^2-x_2^2 = - R^2$ in the
three-dimensional flat space with signature $(+,-,-)$,
implies the $SL(2,R)$ isometry of above metric. 

We will now consider
the dynamics of a single particle
in $AdS_2$ spacetime. Action
$S = - m \int\!ds$ gives the constraint
$g^{\mu\nu}p_{\mu}p_{\nu} = - m^2$, which 
can be solved in the static gauge to obtain 
the Hamiltonian,
\beq
H = - p_0 = \sqrt{\; p^2\, +
\, \frac{m^2R^2}{\sin^2\sigma}\;}  \label{hamiltonian}
\eeq   
Here $p = p_1$ is conjugate to the position coordinate $\sigma$
of the particle. We have used the $AdS_2$ metric
(\ref{metric}).

Due to the $SL(2,R)$ isometry of $AdS_2$ we expect
to realise $SL(2,R)$ on the phase-space,
with Hamiltonian
as one of the generators. 
One can check, using  
the Poisson bracket $\{ \sigma , p \} = 1$, that
$H$ along with $K$ and $J$ defined below,
satisfy the $SL(2,R)$ algebra.
\beq
K\, \equiv\,2 p\,\sin\sigma \;\;\;,\;\;\;
J\, \equiv\,2 \cos\sigma\;\;
\sqrt{\; p^2\, +\, \frac{m^2R^2}{\sin^2\sigma} } 
\label{generators} 
\eeq
\beq
\{ H , K \} = - J \;\; , \;\; 
\{ H , J \} = - K \;\; , \;\;
\{ K , J \} = - 4 H   \label{sl2r}
\eeq

Now, let us note that the group of spatial diffeomorphisms
of $AdS_2$ is generated by the vector fields 
$V_m = 2 \sin{m\sigma}\,\,\frac{\partial}{\partial\sigma}$ ,
where $m$ is a positive integer. 
These spatial diffeos can be
realised on the phase space by 
the functions $K_m$ 
\beq
K_m = 2 p\,\sin{m\sigma} 
\eeq
Above form of
$K_m$, as well as that of $V_m$,
is suggested by the fact that the 
spatial diffeos should leave the
boundary points unaffected. 
The $K_m$ , as expected, satisfy
following relation
\beq
\{K_m , K_n\} = (m-n) K_{m+n} - (m+n) K_{m-n} 
 \label{spacediff}
\eeq
where $m > n$.
Note that $K_1$ is identical to one
of the $SL(2,R)$ generators 
in eqn.(\ref{generators}), $K_1 = K$.

At this point we are led to ask whether
the system we are considering realises
Virasoro algebra, with $H \sim
L_0$ and $K_m \sim
(L_m - L_{-m})$, where $L_n$
are Virasoro generators.
In the following, while discussing
the case of a single particle on $AdS_2$,
we will 
be able to write down
$L_n$ operators
such that $L_0 = \hat{H}$, and the $n \geq 0$
or $n \leq 0$ subset
of the $L_n$ satisfy Virasoro relation.
However, it turns out that the operators
$(L_m - L_{-m})$
do not have the commutation
relations (\ref{spacediff}) expected of spatial diffeo 
generators.
 
It is suggestive   
to define $\Theta_1 ,\Theta_2$ :
\beq
\Theta_1 = \frac{1}{2} ( \, i K_1 \,- \,  \{ K_1\, , \, H \} \,)
\;\;,\;\;
\Theta_2 = \frac{1}{2} 
( \, i K_2 \, - \,\frac{1}{2} \{ K_2 \, , \, H \} \,)  
\eeq
which have the property
\beq
\{ \Theta_1 , H \} = i\, \Theta_1 \;\; , \;\;
\{\Theta_2 , H \} = 2i\, \Theta_2
\eeq
Above expressions hint that the operators corresponding
to $\Theta_1 ,\Theta_2$ may be suitable ansatz for $L_1 ,L_2\,$.
Although for generic values of $(mass)^2$ of the particle
this will not turn out to be correct, in case of 
$(mass)^2 \rightarrow 0$ the operator analogues of 
$\Theta_1 ,\Theta_2$ will indeed give us $L_1 ,L_2$ ,
which will be used to obtain $L_n$ for any $n$.

Next, in order to set up the quantum mechanics of this system,
we first find the Hilbert space where
\beq
H^2 = p^2 + \frac{m^2R^2}{\sin^2\sigma}
\eeq
is represented as a hermitian, positive-definite operator. 
A basis of states for this Hilbert space
is obtained by solving the eigenvalue
problem 
\beq
[\, - \frac{\partial^2}{\partial\sigma^2}\, + 
\, \frac{m^2R^2}{\sin^2\sigma} \, ]\,
\psi = \lambda \, \psi
\eeq
The eigenfunctions can be found to be \cite{perelomov}
\beq
\psi^\alpha_n = (-\sin{\sigma})^{\alpha}\, 
C^{\alpha}_n(\cos\sigma)  \label{basis}
\eeq
where $C^{\alpha}_n$
are Gegenbauer polynomials \cite{grad}.
The corresponding eigenvalues are $\lambda^{\alpha}_n 
= ( n + \alpha )^2 $ where $n = 0, 1, 2, \ldots$ and
$\alpha = (\; 1 + \sqrt{ 1 + 4 m^2R^2 }\;)/2$ .
In order to ensure hermiticity of $H^2$ we must
require $m^2 \, \geq \, -1/(4R^2)$ \cite{perelomov}.
Now, we represent the Hamiltonian
on above Hilbert space by giving its
action on the basis states as 
\beq
\hat{H} \, \psi^\alpha_n = 
(n+\alpha)\, \psi^\alpha_n  \label{hop}
\eeq
Thus the spectrum is independent of mass of the 
particle, except for a constant shift by 
$\alpha(m^2R^2)$. 

For further considerations in this section,
we will first discuss the case of
$(mass)^2 \rightarrow 0$, and then generalise to
$(mass)^2 \neq 0$.

\subsection*{2.1 $(mass)^2 \rightarrow 0$ case}

We will first try to realise $SL(2,R)$ on the
states of single particle system.
Representing $p$ by $-i \frac{\partial}{\partial\sigma}$
on the states gives us 
the operator corresponding to $i K_1$ as
\beq
T_1 = 
2 \sin\sigma\, \frac{\partial}{\partial\sigma} +
\cos\sigma    \label{t1}
\eeq
The second term in above expression is due to the
requirement $T_1^{\dag} = - T_1$.
We require this anti-hermiticity
of $T_1$ because the corresponding
classical observable $i\, K_1$ is pure imaginary.
We have used the standard norm on the states to define
the hermitian conjugate operator.  
Define $L_1, L_{-1}, L_0$ as follows
\beq
L_1 \equiv \frac{1}{2}\, (\,T_1+ [ T_1 , L_0 ] \,) \;\;,\;\;
L_{-1} \equiv - \frac{1}{2}\, (\,T_1- [ T_1 , L_0 ] \,) \;\;,\;\;
L_0 \equiv \hat{H}  \label{l1}
\eeq
Note that $L_1^\dag = L_{-1}$ .
Since $L_0 = \hat{H}$ is defined by explicitly
giving its action on basis states,
eqn.(\ref{hop}), the commutation 
relations of $L_1, L_{-1}, L_0$ are checked 
by considering their action on basis states.
Note that since we are presently taking $(mass)^2 = 0$, $\alpha = 1$
in above expressions for the states and the eigenvalues. 
We find
\beqa
&&L_1 |m\rangle_1 = (m + 1/2) |m-1\rangle_1 \;\;,\;\;\; m \geq 1
\nonumber \\
&&L_{-1} |m\rangle_1 =
(m + 3/2) |m+1\rangle_1    \;\;\;,\;\; m \geq 0
\eeqa
and $L_1|0\rangle_1 = 0$.
We are using $|m\rangle_{\alpha}$ to
denote the normalised basis state
$\psi^\alpha_m$ in equation(\ref{basis}).
Obviously
\beq
[L_1 , L_0] = L_1 \;\; , \;\;   
[L_{-1} , L_0] = - L_{-1} 
\eeq
Furthermore, one obtains $[L_1 , L_{-1}] = 2 L_0$ for action
on the states orthogonal to the ground state.
Surprisingly, on the ground state
\beq
[L_1 , L_{-1}]|0\rangle_1 = (9/4) |0\rangle_1 \neq
2 L_0|0\rangle_1 
\eeq
Next, similar to the case of $i K_1$, 
we represent $i K_2$ by the operator $T_2$ 
\beq
T_2 = 2 ( \sin{2\sigma}\, \frac{\partial}{\partial\sigma} +
\cos{2\sigma} )\;\;\;\; , \;\;\;\; T_2^{\dag} = - T_2 
\eeq
Define $L_2$ and $L_{-2}$ as 
\beq
L_2 = (\, [\, T_2\, ,\, L_0 \,]\, +
\, 2 \,T_2 \,)/4    \;\;\;\; , \;\;\;\;
L_{-2} = (\, [\, T_2\, , \, L_0 \,]
\, - \, 2\, T_2\, )/4   \;\;\;\;,\;\;\;\;
L_2^\dag = L_{-2}
\label{l2}
\eeq
We obtain
\beqa
&&L_2 |m\rangle_1 = m |m-2\rangle_1 \;\;\;\;\{\; m \geq 2\;\}
\nonumber \\
&&L_2 |m\rangle_1 = 0   \;\;\;\;     \{\; m < 2\;\}  \nonumber \\
&&      \nonumber \\
&&L_{-2} |m\rangle_1 =
(m + 2) |m+2\rangle_1    \;\;\;\;  \{\; m \geq 0 \;\}
\eeqa
Notice that the relation $[\,L_1,L_{-2}] = 3 L_{-1}$
is satisfied on all the states orthogonal to the ground
state.
Further, we define $L_n$ for $n \geq 3$ by
\beq
(n-2) \, L_n \, = \, [L_{n-1} \, , \, L_1]
\eeq
and $L_{-n} \equiv L_n^{\dag}$.
Using above definitions one can obtain following
expressions for the action of $L_n$ and $L_{-n}$ 
on the basis states. For $n \geq 1$ 
\beqa
&&L_n |m\rangle_1 = (m - n/2 + 1) |m-n\rangle_1 \;\;\;\;\{\; m \geq n\;\}
\nonumber \\
&&L_n |m\rangle_1 = 0   \;\;\;\;     \{\; m < n\;\}  \nonumber \\
&&      \nonumber \\
&&L_{-n} |m\rangle_1 =
(m + n/2 + 1) |m+n\rangle_1    \;\;\;\;  \{\; m \geq 0 \;\}
\eeqa
Remarkably, the 
$L_n$ for $n \geq 0$,
satisfy semi-infinite
Virasoro algebra, as can be easily checked using above
equations. 
\beq
[\, L_n\, ,\, L_m \,] \, = 
\, (n-m)\, L_{n+m} 
\eeq
Due to $L_{-n} = L_n^\dag$, the $L_n$ for $n \leq 0$
also satisfy above semi-infinite Virasoro algebra.

\subsection*{2.2 $(mass)^2 \neq 0$ case}

Let us first consider realisation of $SL(2,R)$.
We may naively continue to take the 
expression for $T_1$ of the
form (\ref{t1}), 
and use that to define $L_{\pm 1} , L_0$ as in (\ref{l1}).
However, with 
$L_{\pm 1} , L_0$ defined in this way the
$SL(2,R)$ relations are not satisfied, not only on the
ground state as it happened in case of $\alpha = 1$, but generically
on any state. It is possible to find an appropriate modification
of $T_1$ so that $SL(2,R)$ is realised in a way similar to 
the case of $\alpha = 1$. 
We consider following realisation of $T_1$
which by eqn.(\ref{l1}) gives $L_{\pm 1}$.
As before $L_0 = \hat{H}$. 
\beqa
&&T_1 |0\rangle_{\alpha} = - ( \alpha + 1/2 ) |1\rangle_{\alpha} \nonumber
\\
&&T_1 |m\rangle_{\alpha} = ( \alpha + m - 1/2 ) |m-1\rangle_{\alpha}
- ( \alpha + m + 1/2 ) |m+1\rangle_{\alpha} \;\;\;\; \{\; m \geq 1\;\}
\eeqa
One can check that $T_1^{\dag} = -\,T_1$ and 
\beq
L_1 |m\rangle_{\alpha} = (\alpha + m - 1/2) |m-1\rangle_{\alpha}
\;\; , \;\; \{\, m \geq 1 \,\} \;\;\;,\;\;\;
L_1 |0\rangle_{\alpha} = 0 
\eeq
\beq
L_{-1} |m\rangle_{\alpha} = (\alpha + m + 1/2) |m+1\rangle_{\alpha}
\;\; , \;\; \{\; m \geq 0 \;\} 
\eeq
Above equations give: 
$[ L_{-1} , L_0 ] = - L_{-1}\; ,\; [ L_1 , L_0 ] =  L_1$.
Similar to the $\alpha=1$ case,
the relation $[ L_{-1} , L_1 ] = -2 L_0$ holds 
on states orthogonal to ground state. On the ground state
we get
\beq
[ L_{-1} , L_1 ]|0\rangle_{\alpha} = -(\alpha+1/2)^2 |0\rangle_{\alpha}
\neq -2 L_0|0\rangle_{\alpha}
\eeq
Note that for $\alpha = 1/2$, which is the case of
$m^2 = -\frac{1}{4R^2}$, the
relation $[ L_{-1} , L_1 ] = -2 L_0$ is satisfied on all
states.

Next, we will find $L_{-2}$ by requiring that it satisfies following
conditions
\beq
[\,L_0\,,\,L_{-2}\,]\;=\;2L_{-2}
\eeq
\beq
[\,L_1\,,\,L_{-2}\,]|\psi\rangle_{\alpha} \;
=\;3L_{-1}|\psi\rangle_{\alpha}
\eeq
where $|\psi\rangle_{\alpha}$ is any state orthogonal
to the ground state $|0\rangle_{\alpha}\,$.
Above relations were satisfied
by $L_{-2}$ in $\alpha=1$ case.
First condition gives
\beq
L_{-2}|m\rangle_{\alpha} = a_m\,|m+2\rangle_{\alpha} \;\;,\;\; m \geq 0 
\eeq
and second condition gives a recursion relation for $a_m$s,
which determines all the $a_m$ in terms of $a_0$.
Further, we define $L_{-n}$, for $n \geq 3$, by
\beq
-(n-2) \, L_{-n} \, = \, [L_{-(n-1)} \, , \, L_{-1}]
\eeq
Thus, we have obtained an ansatz for $L_{-n}$ in terms of an unknown
parameter $a_0$ which will be fixed by requiring
\beq
[ L_{-4} , L_{-1} ]|0\rangle_\alpha = 3 [ L_{-3} , L_{-2} ]|0\rangle_\alpha
\eeq
Finally we obtain, for $n \geq 1$,
\beqa
&&L_n |m\rangle_{\alpha} =  (m - n/2 + \alpha ) |m-n\rangle_{\alpha}
\;\;\;\;\{\; m \geq n\;\}
\nonumber \\
&&L_n |m\rangle_{\alpha} = 0   \;\;\;\;     \{\; m < n\;\}  \nonumber \\
&&      \nonumber \\
&&L_{-n} |m\rangle_{\alpha} =
(m + n/2 + \alpha ) |m+n\rangle_{\alpha}    \;\;\;\;  \{\; m \geq 0\;\}
\label{ln}
\eeqa
where $L_n \equiv L_{-n}^\dag$.
Similar to the case of $\alpha = 1$, we find
\beq
[\, L_n\, ,\, L_m \,] \, =
\, (n-m)\, L_{n+m} \;\;,\;\; \{\, n, m\, \geq\, 0 \} \label{semivir}
\eeq
for general value of $\alpha\,$.

\subsection*{3. $N$ Particles on $AdS_2$}

So far we have shown that Hilbert space of a single 
particle on $AdS_2$ realises semi-infinite Virasoro algebra.
The aim of this section is to
demonstrate that when we consider a multi-particle
system on $AdS_2$, we obtain a structure that is reminisent
of $c = 1$ Virasoro algebra.

Introducing interaction between particles on $AdS_2$ appears
to be complicated because of the requirement of
$SL(2,R)$ symmetry that follows from 
$AdS_2$ isometry. 
We will restrict our discussion 
here to the case of non-interacting 
particles. However, as we will now argue, 
considering this system to be a 
limit of $SL(2,R)$ compatible
interacting system where the interaction
strength is gradually taken to zero 
suggests that we must treat these
identical particles as fermions.

Let us first note that the energy spectrum of a single
particle on $AdS_2$, eqn.(\ref{hop}), agrees exactly with that
of the system described by following hamiltonian 
\beq
{\cal H} = \frac{p^2}{4} +
\frac{g}{4 q^2} + 
\frac{q^2}{4}   \label{mdff} 
\eeq
if we choose $g = (3 + 16 m^2R^2)/4\,$. The spectrum of above 
hamiltonian \cite{dff} is $E_n = (n + \beta)$,
where $\beta = (1+\sqrt{g + \frac{1}{4}\,})/2$
and $n = 0,1,..\,$.
Furthermore $\cal H$, along with $\cal K$ and $\cal J$,
defined below, form $SL(2,R)$ similar to $H, K, J$ in 
eqn.(\ref{sl2r}).
\beq
{\cal K} = pq \;\;,\;\;
{\cal J} = \frac{p^2}{2} +
\frac{g}{2 q^2} - 
\frac{q^2}{2} 
\eeq
\beq
\{ {\cal H} , {\cal K} \} = - {\cal J} \;\; , \;\;
\{ {\cal H} , {\cal J} \} = - {\cal K} \;\; , \;\;
\{ {\cal K} , {\cal J} \} = - 4 {\cal H}   
\eeq
Since the spectrum and symmetry of above single particle 
system agree with that of a particle on $AdS_2$,
multiparticle generalisation of both systems,
such that $SL(2,R)$ symmetry is satisfied,
should have similar properties.
Now, consider $N-$particles generalisation 
of above system (\ref{mdff}), allowing for interaction such that
${\cal H}_N, {\cal K}_N, {\cal J}_N$ form $SL(2,R)$ .
One finds that
\beq
{\cal H}_N = \frac{1}{4} \sum_{i=1}^{N} p_i^2  +
\sum_{i=1}^N \, \frac{g}{4 q_i^2} +
\sum_{i < j}  \frac{\lambda}{(q_i - q_j)^2} +
\frac{1}{4} \sum_{i=1}^{N} q_i^2    \label{mcalo} 
\eeq
\beq
{\cal K}_N = \sum_{i=1}^N p_iq_i
\;\;\;\;,\;\;\;\;
{\cal J}_N = \frac{1}{2} \sum_{i=1}^{N} p_i^2  +
\sum_{i=1}^N \, \frac{g}{2 q_i^2} +
\sum_{i < j}  \frac{2 \lambda}{(q_i - q_j)^2}  - 
\frac{1}{2} \sum_{i=1}^{N} q_i^2
\eeq
It is clear from the interaction term in ${\cal H}_N$
that in the limit $\lambda \rightarrow 0$, the short
interparticle distance behaviour of wavefunctions is 
$\psi_N (q_i , q_{i+1}) \sim (q_i - q_{i+1})$.
This implies that in $\lambda \rightarrow 0$ limit
the system is equivalent to that of non-interacting 
fermions. Therefore the non-interacting particles on
$AdS_2$ should also be treated as fermions.

Coming back to our original system
we have the Hamiltonian 
\beq
H_N = \sum_{i=1}^N \,
\sqrt{ \, p_i^2 \, + \, \frac{m^2R^2}{\sin^2\sigma_i} \,}
\eeq
For reasons to be clear in the following, it will be
preferable to take $L_0$ in the $N$ particles case to be
shifted with respect to the Hamiltonian
\beq
L_0 = H_N\;+\; (\,\alpha - 1/2\,)^2\;/2 
\eeq
We will define the $L_n$ for $N$-particles system to be 
the sum of single particle $L_n$, which is denoted
below by $L_n^i$ for the $i$th particle.
\beq
L_n = \sum_{i=1}^N \; L_n^i 
\eeq
The N-particle basis states are
\beq
|n_1,n_2,\ldots,n_N\rangle_{\alpha} = det[M_{ij}] \;\;,\;\;
M_{ij} = \psi_{n_i}^{\alpha}(\sigma_j) \label{nbasis}
\eeq
and the ground state $|vac\rangle_{\alpha}$ is characterised by
$n_i = (i-1)$ in above formula.

Let us first note that due to eqn.(\ref{semivir}) it is obvious 
that for
$\;n, m\;\geq\;0$, or $n, m\;\leq\;0$, 
\beq
[\,L_n\,,\,L_m\,]\,=\,(n-m)\,L_{n+m} \label{nsemivir} 
\eeq
Furthermore, for $n \geq 1$
\beq
L_n|vac\rangle_{\alpha} = 0 
\eeq
One can check that $L_1, L_{-1}, L_0$ satisfy the
$SL(2,R)$ relations on any state of excitation 
energy less than $N$.

Consider now $[\,L_n\,,\,L_{-m}\,]|vac\rangle_{\alpha}\,$
for $n,m\,\geq \,1\,$. 
\beq
L_{-m}|vac\rangle_{\alpha} =
\sum_{i = i_0}^{m-1} (i+N-\frac{m}{2}+\alpha)
|0,1,\ldots,N+i-m-1,N+i,N+i-m+1,\ldots\rangle_{\alpha}
\eeq
where $i_0=0$ if $N \geq m \geq 1$, and $i_0=(m-N)$
if $m > N$.
Now, for $n>m$ 
\beq
L_n|0,1,\ldots,N+i-m-1,N+i,N+i-m+1,\ldots\rangle_{\alpha}
= 0
\eeq
Therefore,
for $n\,>\,m\,\geq\,1$,
$[\,L_n\,,\,L_{-m}]|vac\rangle_{\alpha}\,
=\,L_n\,L_{-m}|vac\rangle_\alpha\,=
\,0\,=\,(n+m)\,L_{n-m}|vac\rangle_\alpha\,$.
\beq
L_m|0,1,\ldots,N+i-m-1,N+i,N+i-m+1,\ldots\rangle_{\alpha}
= (N+i-m/2+\alpha) |vac\rangle_{\alpha}
\eeq
Therefore, for $N \geq m \geq 1$
\beq
[L_m\,,\,L_{-m}]|vac\rangle_{\alpha}\,=\,
\{\,\sum_{i = 0}^{m-1} \, (N+i-m/2+\alpha)^2\,\}\,\,
|vac\rangle_{\alpha}\,=\,
2 m\, L_0|vac\rangle_\alpha \,+\, 
\frac{1}{12} (m^3 - m)|vac\rangle_\alpha
\eeq
For $n\,<\,m\,\leq\,N$
\beqa
&&[\,L_n,L_{-m}\,]
|vac\rangle_\alpha\,=\,L_n L_{-m}|vac\rangle_{\alpha} = \nonumber \\
&&\sum_{i=n}^{m-1}\! (i\!+\!N\!-\!\frac{m}{2}\!+\!\alpha)
(N\!+\!i\!-\!\frac{n}{2}\!+\!\alpha)
|\ldots,N\!+\!i\!-\!m\!-\!1,
N\!+\!i\!-\!n,N\!+\!i\!-\!m\!+\!1,\ldots\rangle_{\alpha}
\nonumber \\
&&-\!\sum_{i=0}^{m-n-1}\! (i\!+\!N\!-\!\frac{m}{2}\!+\!\alpha)
(N\!+\!i\!-\!m\!+\!\frac{n}{2}\!+\!\alpha)
|..,N\!+\!i\!-\!m\!+\!n\!-\!1,
N\!+\!i,N\!+\!i\!-\!m\!+\!n\!+\!1,..\rangle_{\alpha} \nonumber \\
&&= (n+m)\! \sum_{i=0}^{m-n-1}\! (i\!+\!N\!-\!\frac{m-n}{2}\!+\!\alpha)
|\ldots,N\!+\!i\!-\!m\!+\!n\!-\!1,
N\!+\!i,N\!+\!i\!-\!m\!+\!n\!+\!1,\ldots\rangle_{\alpha} \nonumber \\
&&= \,(n+m) \; L_{n-m}|vac\rangle_{\alpha}
\eeqa
Thus we have found that, for $N \geq n \geq 1$ and $-N \leq m \leq -1$
\beq
\{\,[\,L_n , L_m\,]\,-\, (n-m)\, L_{n+m} \,-\,
\frac{1}{12} \, (n^3 - n) \delta_{n+m , 0}\,\}\,|vac\rangle_\alpha
\,=\,0
\eeq
Similarly it is straightforward to check that
\beq
\{\,[\,L_n , L_m\,]\,-\, (n-m)\, L_{n+m} \,-\,
\frac{1}{12} \, (n^3 - n) \delta_{n+m , 0}\,\}\,L_{-i}|vac\rangle_\alpha
\,=\,0
\eeq
where $(N-i) \geq n \geq 1$ and $-(N-i) \leq m \leq -1$.
Above relations, along with eqn.(\ref{nsemivir}) are
reminisent of $c = 1$ Virasoro algebra, although in above we
have a $(N-i)$ dependent
cut-off on the set of $L_n$s for which the 
Virasoro-like relations hold.
In general we expect that on a state with excitation
energy $\epsilon$, the $L_n$s,
$(N-\epsilon) \geq n \geq -(N-\epsilon)$,
have $c = 1$ Virasoro-like relation.

\subsection*{4. Conclusion}

We have restricted our study to
the case of non-interacting particles on $AdS_2$.
It is important to see whether
our discussion generalises to
interacting system. In this context
one may study the model 
described by (\ref{mcalo}), as our discussion
in last section suggests that this model
is closely related to a multiparticle system
on $AdS_2$. The system described by (\ref{mcalo}) is a variant
of Calogero model \cite{calo}. It is interesting to note that a
supersymmetric variant of Calogero model, but without the
$g/q_i^2$ term in (\ref{mcalo}), 
was suggested \cite{gt} to
describe microscopic degrees of freedom of certain extremal
black-holes.

A surprising aspect of our analysis of a
single particle dynamics on $AdS_2$ is that $SL(2,R)$ is
not exactly realised\footnote{Except for 
$\alpha = 1/2$, where it is exactly realised
on all the states.}.   
In the model \cite{dff}
of (\ref{mdff}), which has the same spectrum,
one can write the operators
$L_{\pm}\,, L_0$ that generate $SL(2,R)$.
\beq
L_0 = 
\frac{p^2}{4} +
\frac{g}{4 q^2}
+ \frac{q^2}{4}
\;\;,\;\; 
L_{\pm} = 
\frac{p^2}{4} +
\frac{g}{4 q^2} 
- \frac{q^2}{4} \mp   
\frac{i}{4} (pq + qp) 
\eeq
It will be interesting to find out whether
the algebra of $L_{\pm}, L_0$ 
enlarges to a Virasoro-algebra like structure,
when multiparticle system is considered,
in a way similar to our discussion in this paper.

\subsection*{Acknowledgements}

I would like to thank K.S. Narain for useful discussions
and for bringing ref.\cite{gt} to my attention.
I would also like to thank the High Energy Theory Group of
ASICTP, Trieste, for their hospitality while this 
work was being completed.

\end{document}